\documentclass[conference]{IEEEtran}
\IEEEoverridecommandlockouts
\usepackage[nocompress]{cite}

\usepackage{url}
\usepackage{amssymb}
\usepackage{pifont}
\newcommand{\cmark}{\ding{51}}%
\newcommand{\xmark}{\ding{55}}%
\usepackage{amsmath,amssymb,amsfonts}
\usepackage{algorithmic}
\usepackage{graphicx}
\usepackage{textcomp}
\usepackage{xcolor}
\usepackage{comment}
\usepackage{multirow}

\def\BibTeX{{\rm B\kern-.05em{\sc i\kern-.025em b}\kern-.08em
    T\kern-.1667em\lower.7ex\hbox{E}\kern-.125emX}}
\usepackage{pifont}
\let\oldding\ding
\renewcommand{\ding}[2][1]{\scalebox{#1}{\oldding{#2}}}%

\begin{document}

\title{ELMO2EDS: Transforming Educational Credentials into Self-Sovereign Identity Paradigm
\thanks{This paper has been realized via funding of the IDunion project by
German Ministry of Economic Affairs and Energy (BMWK), organized by
DLR Projektträger}
}

\author{\IEEEauthorblockN{Patrick Herbke}
\IEEEauthorblockA{\textit{Service-centric Networking} \\
\textit{Technische Universität Berlin}\\
Berlin, Germany \\
p.herbke@tu-berlin.de}
\and
\IEEEauthorblockN{Hakan Yildiz}
\IEEEauthorblockA{\textit{Service-centric Networking} \\
\textit{Technische Universität Berlin}\\
Berlin, Germany \\
hakan.yildiz@tu-berlin.de}
}

\maketitle

\begin{abstract}
\textbf{Digital credentials in education make it easier for students to apply for a course of study, a new job, or change a higher education institute. Academic networks, such as EMREX, support the exchange of digital credentials between students and education institutes. Students can fetch results from one educational institute and apply for a course of study at another educational institute. Digital signatures of the issuing institution can verify the authenticity of digital credentials. Each institution must provide the integration of EMREX using its identity management system. In this paper, we investigate how digital credentials can be integrated into the Self-Sovereign Identity ecosystem to overcome the known issues of academic networks. We examine known issues such as the authentication of students. Self-Sovereign Identity is a paradigm that gives individuals control of their digital identities. Based on our findings, we propose ELMO2EDS, a solution that 1) converts digital credentials from EMREX to a suitable Self-Sovereign Identy data format, 2) enables authenticating a student, and 3) enables issuing, storing, and verification of achieved study.}
\end{abstract}

\begin{IEEEkeywords}
Lifelong learning, education, e-learning, learning achievements, Self-Sovereign Identity, SSI, verifiable credentials, digital identity, semantic, syntactic, Europass, EMREX, ELM, ELMO, EBSI, EBP, EDCI, EWP
\end{IEEEkeywords}

\section{Introduction}
Paper-based education credentials usually contain the signature of issuing person(s) and the affiliated institutions' stamp to verify the credentials' authenticity. Digital education credentials in school and higher education institutions (HEIs) have the same requirements as paper-based credentials in terms of their verifiability for authenticity~\cite{kontzinos2020using}. 

Digital credentials (DCs) are an emerging trend in education and the labor market, as DCs facilitates holding, issuing, and verifying of learning achievements' authenticity~\cite{komljenovic2021rise}. The main components of DCs are information (claims) about the student (holder), educational institution (issuer), and learning achievements. Learning achievements of DCs are a digital representation of a physical school and HEI certificate (diploma) or a digital transcript of records~\cite{lemoine2015micro}. Educational networks (ENs) and software applications support exchanging, signing, and verifying of DCs between students, HEIs, and companies~\cite{mincer2017emrex, wagenaar2019reform}. At the same time, semantic and syntactic data standards for DCs evolve~\cite{goger2022learning, bohlinger2019ten, silvestru2018isced}. 

\begin{figure}[t]
  \centering
  \includegraphics[width=0.463\textwidth]{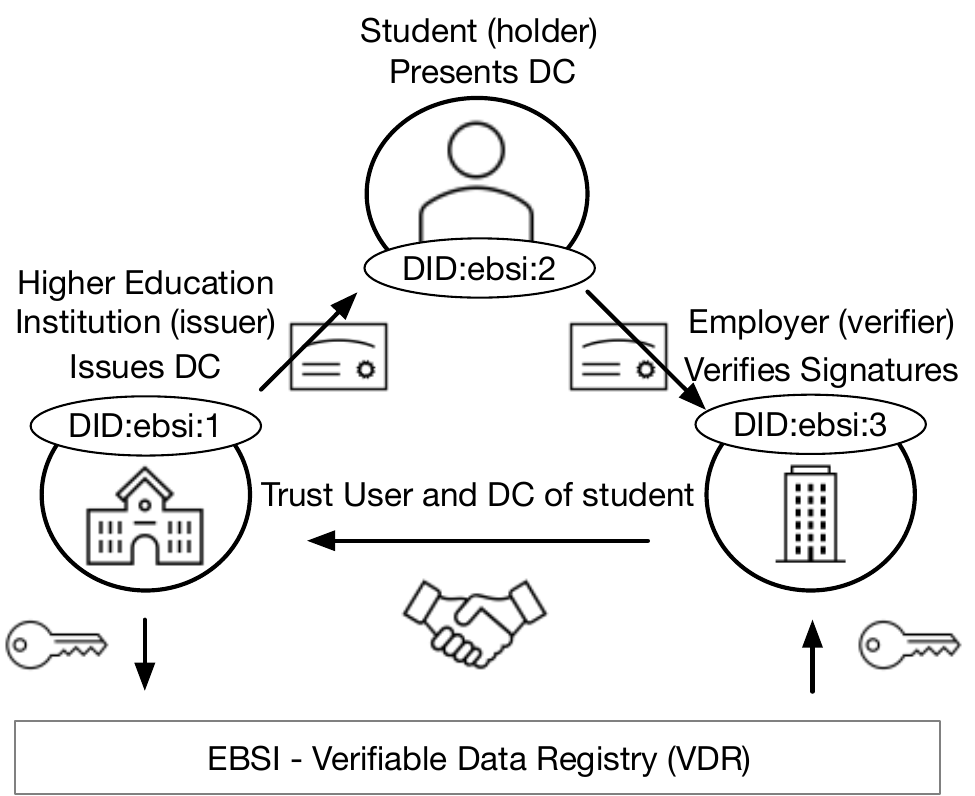}
  \caption{Design of Self-Soverein Identity in education. Issuer, issues DCs to holder. Holder presents verifiable DCs to verifier. Verifier verifies public key of issuer and holder with VDR of EBSI.}
  \label{fig:ssitrust}
\end{figure}

With ENs, students can request and recognize academic achievements from other HEIs. Furthermore, students can apply online with a DC for a study or job. A signed DC enables companies and HEI to automatically verify DCs' authenticity with the issuers' digital signature~\cite{strack2019eidas}. 

Our paper proposes a solution based on EMREX, a European-used educational network. EMREX utilizes public-key infrastructure to issue and proof the authenticity of DCs~\cite{emrex_community_emrex_2022, chokhani2003internet}. EMREX is easily adoptable by HEIs and is not based on blockchain technology. HEIs can fulfill the following challenges with the help of EMREX plugins. Main challenges~\cite{emrextech} described by EMREX are:~\ding[1.2]{192} Authenticating a student,~\ding[1.2]{193} fetching results for a student, and~\ding[1.2]{194} storing results for a student. One of the main challenges addressed in this paper is verifying the authenticity of students. In EMREX, linking digital identities to students is not supported by default. Linking students means being able to verify students' identities digitally. The authentication of students is necessary for accessing DCs and verifying students' identity of DCs. It is a challenge in EMREX because every institution has to connect its technical identity solutions to authenticate students, and student identity within DCs is not supported per default in EMREX.

In the next generation of education, students will be empowered to have self-sovereignty over their data. Figure~\ref{fig:ssitrust} shows the principle of Self-Sovereign Identity (SSI) by Allen\cite{allen_path_2022}. An issuer issues a verifiable DC to a holder. The holder stores the DC in a personal wallet. Then the holder presents the DC to a verifier. The verifier can verify the presented DC through the verifiable data registry (VDR)\footnote{Verifiable data registry: https://www.w3.org/TR/vc-data-model/}. Detailed information on SSI is described in Section II. SSI offers student authentication and verification of the authenticity of \textit{both} the issuer and the holder. It is also an evolving trend in the digitalization of education~\cite{ocheja2022visualization, arndt2020blockchain, deenmahomed2021future, turkanovic2020signing, gish2021micro}.

As a contribution to the development of next-generation ENs, we propose ELMO2EDS\footnote{We provide the source code of ELMO2EDS at: https://github.com/TUB-SSI-Edu/ELMO2EDS}. ELMO2EDS supports the transformation of EMREX exchanged DCs into European Blockchain Service Infrastructure (EBSI). Transformations of EMREX DCs to other technology stacks than EBSI are feasible and described at the end of this paper. EBSI aims to fulfill the SSI paradigm proposed by Allen~\cite{allen_path_2022}. 

This paper presents an approach that converts EMREX exchanged DCs~\cite{emrextech} in EBSI compliant data format~\cite{ebsi_2022}. The main contributions of this paper are:

\begin{itemize}
    \item Reviewing educational data standards
    \item Implementing an application that maps and transforms ELMO-based credentials to EBSI compliant credential format
    \item Discussing limitations of the proposed method
\end{itemize}

The remainder of this paper is organized as follows: Section~\ref{sec:backgr} gives background information on EMREX, SSI, and EBSI. Section~\ref{sec:relwork} surveys ENs and applications. In Section~\ref{sec:des} the design and requirements of our approach are described in detail. Section~\ref{sec:impl} gives an overview of our implementation, followed by Section~\ref{sec:eval}, limitations of our implementation, and methods proposed in this paper. The research results are concluded in Section~\ref{sec:con} followed by an outlook in Section~\ref{sec:out}.

\section{Background}
\label{sec:backgr}

\subsection{EMREX}
EMREX is an Erasmus+ project for exchanging digital credentials between students and HEIs, for example, study or exam results. EMREX consists of the following components\footnote{EMREX Architecture: https://github.com/emrex-eu/standard}: The EMREX client (EMC), EMREX Data Access Points (EMP), and the EMREX registry. The EMREX registry is part of Erasmus-without-paper (EWP)~\cite{hokkanen2019mobility}.

The EMREX system is an application that students can use to request data stored by an HEI via an EMP. Once a student requests data from another HEI, the EMC connects to the EMP of the requested HEI. An EMP contains information about the name, country, contact URL, and public key of participating HEI. Any EMC implemented in the student information system of an HEI can contact any EMP by establishing a connection with the contact URL of the EMP. The connection enables students to log in using their access data from the requested HEI and transfer data. The requested data is retrieved via the EMP using the EMREX protocol~\cite{berbecaru2019providing}. 

ELMO is used in the EMREX network for exchanging DCs. It enables signing and verifying of DCs with XML signature\footnote{XML-Signature: https://www.w3.org/TR/xmldsig-core/}. Requested DC by a student is signed by the issuing HEI with their private key and can be verified for authenticity by other associated HEI. The verifying HEI uses the public key of the issuing HEI in the EMP. 

The EWP registry lists all EMPs, separated for each country. It contains the access data to the EMPs of associated HEIs and is, therefore, a central component for participating HEIs. 

To join EMREX as an EMC, an HEI must adapt its identity management system (e.g., IdM, SiS) to EMREX plugins. A further step for HEI is creating a private and public key with a certified authority of the corresponding country. After creating the certificate, the generated public key is stored in the EMP and used to verify students' data.

\subsection{Self-Sovereign Identity}
SSI is a recent identity paradigm that empowers identity subjects (holders) to have complete control and ownership of their identities~\cite{tobin2016inevitable}. Similar to paper-based credentials, self-sovereign identities can exist without the issuing party and, therefore, are not locked in with the issuing parties. The vision of SSI is a user-centric and decentralized approach to digital identities. Digital identities can exist even if the issuing party does not anymore. The realization of this decentralized approach requires the creation of trust and the source of truth without relying on a public-key infrastructure. Instead, commonly shared, highly available, and tamper-proof VDRs governed by multiple organizations and stakeholders play a vital role in \mbox{establishing trust}.

In SSI, every entity has a decentralized identifier (DID). A DID is an identifier that follows the syntax of Uniform Resource Identifiers. It is globally unique, persistent, and entirely under the control of the entity it represents. Through a DID method and resolve instruction in it, a DID resolves into a DID document that contains information such as public keys and service endpoints related to the DID. DIDs and DID documents can be stored in VDRs, which is a source of truth due to the tamper-proof property that comes with them. With this constellation, a DID Document of a DID can be compared to a digital certificate like X.509\footnote{X.509 Certificate: https://datatracker.ietf.org/doc/html/rfc5280}. Hence, it guarantees that a public key belongs to a certain DID Subject and not to an entity that pretends to be this DID subject. W3C~\cite{did-core} standardizes DIDs and DID documents.

A digital identity that an issuer fabricates is called a ntial (VC). A VC is a tamper-proof container that contains statements about the subject. The origin of the claims is cryptographically verifiable. The verifiability comes with a signature created by a private key belonging to a public key. DIDs relate to a specific DID Document stored in the VDR. A DID Document contains the public key of the issuer. Verifier uses the public key to check the authenticity of the issued credential. VCs are tamper-resistant because if the content of the credential is tampered with, the issuers' signature does not fit the tampered content. VCs are bound to the identity holder. One binding mechanism involves holder DID being written in the VC. The holder then presents a derivation of the VC called a verifiable presentation. A verifiable presentation is one-time use only and contains proofs created with the holders' private key. The public key of the holder, stored in the DID document, relates to the holders' DID and is used to verify presented credential claims.





\subsection{European Blockchain Service Infrastructute}
The European Blockchain Service Infrastructure (EBSI) is a joint initiative of the European Commission and the European Blockchain Partnership (EBP). In the EBP, European countries work together on blockchain-based public services for EU citizens and organizations~\cite{olnes2021blockchain, 9206057}. HEIs, organizations, and research institutions of the EBP member countries design and implement technical requirements and political regulations for EBSI, especially regarding cross-border use cases~\cite{buchanan2018building}.

EBSI aims to create digital trust in the EU. Therefore development is oriented to EU regulation No. 910/2014, electronic identification and trust services (eIDAS). A description of EBSI gives Ølnes~\cite{grech2021blockchain} and Tan et al.~\cite{tan2022blockchain}. Schwalm et al.~\cite{Schwalm2022} describe the challenges and opportunities of eIDAS 2.0 and the requirements of (European) standardization. 

EBSI is a public-permissioned blockchain based on Proof of Authority consensus~\cite{lykidis2021use}. The APIs of EBSI provide the functionality to interact with Hyperledger Besu and Fabric. Hyperledger Besu is an Ethereum Client running on Ethereum Mainnet. Hyperledger Fabric is a permissioned distributed ledger technology platform. EBSI plans to use the European Learning Model (ELM) as the data model for educational data, as described by Fridell et al.~\cite{elmTor}. 

This paper contributes to EBSIs' diploma use case as our implementation links the educational network EMREX and EBSI. The following section presents related work of digitization in education.

\section{Related Work}
\label{sec:relwork}
Educational networks are becoming increasingly important as new forms of teaching evolve. Traditional forms of teaching confront digital educational credentials and modern e-learning systems~\cite{brands2000rethinking, ahel2020opportunities, habibi2020digitalization}. 
\subsection{Educational Platforms}
The number of online educational platforms is increasing in lower and HEIs~\cite{e-learning-plattform_2022}. There is also a wide range of education-related platforms for different domains. For example, Europass enables citizens to create a Europe-wide verifiable curriculum vitae for job applications\cite{europass_2022}. Europass also offers citizens a portfolio for digital credentials. Bacharach et al.~\cite{bacharach2021progress} present Status and future perspectives of digitization of higher education and describe the relation of EMREX to external standards and co-existing ENs. At the same time of evolving ENs, the number of research areas for blockchain-based applications in education is increasing~\cite{grech2017blockchain, turkanovic2018eductx}. 

Yildiz et al.~\cite{yildiz2021connecting} developed a solution to connect SSI with Federated and User-centric Identities. In more detail, they developed an application enabling authentication via SSI-based credentials. With these credentials, students can log in into an HEI portal. DFN~\cite{dfn_dokumentation_2022} and SWITCH~\cite{hassenstein_swiss_nodate} are working on an educational identifier for authenticating students.

Yumna et al.~\cite{yumna2019use} give an overview of blockchain-based educational applications. They summarize research questions and introduce state-of-the-art applications in education. Turkanović et al.~\cite{8247166} present eduCTX. In eduCTX, students retrieve tokens representing credits for achievements, such as ECTS credits. The next paragraph presents related work of EBSI.

\subsection{Transformation}
Abraham et al.~\cite{abraham2018qualified} shows a method to transform an electronic identifier from existing systems to SSI. Fedrecheski et al.~\cite{low-overhead_approach} propose transforming the internet of things domain to SSI with metadata serialization. Garzon et al.~\cite{6gssi} introduces opportunities and potential benefits of using SSI for cross-actor and privacy-preserving identity and key management in the following mobile network generation 6G. Another significant contribution to sustainability gives Berg et al.~\cite{berg2022overcoming}. They present Digital Product Passports that revolutionize plastic value chains with self-sovereign identity. 

\section{Design}
\label{sec:des}
This section describes design decisions for the development of ELMO2EDS. The design decisions are based on comparing data schemas and signature procedures between EMREX and EBSI. The motivation for the development arises from ample discussions in the consortiums \textit{IDunion}\footnote{IDunion: https://idunion.org/}, \textit{Network Digital Evidences}\footnote{Network Digital Evidences(DE): http://netzwerkdigitalenachweise.de} and \textit{The School Network}\footnote{The School Network(DE): https://das-schulnetzwerk.de/}. 

In this paper, one challenge of EMREX is tackled, the authentication of students\footnote{EMREX technical guide: https://bit.ly/3bxjieI}, as it is not included by default. ELMO2EDS is an application that converts EMREX DCs (upper secondary school, transcript of records) to EBSI diploma schema (EDS)\footnote{EBSI diploma schema: https://bit.ly/3HDG2Wr}. EDS is similar to ELMO, a modified version of ELM, and available for research as of 6/2022.

The proposed converter gets an ELMO XML\footnote{ELMO Schema: https://github.com/emrex-eu/elmo-schemas} credential as input and outputs an EBSI compliant JSON-LD\footnote{JSON for Linking Data: https://json-ld.org/} file. We chose the EBSI diploma schema as the target schema for the output file. The EBSI enables members to access blockchain infrastructure via APIs for trustworthy processes. These processes are between institutions and citizens in the European Union and beyond~\cite{allessie2019blockchain}. Furthermore, EDS enables digital signatures for issuers \textit{and} holders for DCs and aims to realize cross-border use cases. 
\begin{table}[!t]
\caption{External standards used in ELMO and EDS}
\vspace{-5.2mm}
\begin{center}
\begin{tabular}{|l|l|l|}
\hline
\multirow{2}{*}{\textbf{Claims}} & \multicolumn{2}{c|}{\textbf{External Standards}} \\
{} & \textbf{ELMO} & \textbf{EDS}\\
\hline
\textit{Holder (student):} & &\\
citizenship & ISO 3166-1-alpha-2 & W3C VC data model \\
identifier & National \textit{or} European & DID, LEI, EORI\\ 
& (student) identifier & SEED, SIC, VAT \\ 
currentAddress & EWP Address & W3C VC data model \\
gender & ISO/IEC 5218 & W3C VC data model \\
\hline
\textit{Issuer (HEI):} & & \\
country & ISO 3166-1-alpha-2 & W3C VC data model \\
identifier & PIC, EWP or SCHAC & DID, LEI, EORI \\
& & SEED, SIC, VAT \\
currentAddress & EWP Address & W3C VC data model\\
\hline
\textit{Learning Opportunity:} & &\\
gradingScheme & ECTS & \cmark \\
degreeProgramLevel & EQF, NQF & \cmark \\
iscedCode & ISCED-F or EWP LOI & ISCED-F\\
languageOfInstruction & ISO 639-1 & ESCO \\
& & Classification \\ 
\hline
\end{tabular}
\label{tab1}
\end{center}
\end{table}

The tasks of developing ELMO2EDS consist of the~\ding[1.2]{192} analysis of educational standards,~\ding[1.2]{193} converting data from ELMO to EDS,~\ding[1.2]{194} extension of converted data with decentralized identifiers. 

For developing ELMO2EDS, we analyzed the data schemas of ELMO and EDS. ELMO and EDS schemas differ in issuers' and holders' identifiers. In EMREX, student identifiers of DCs are bound to a country or institution. The disadvantage of national student identifiers is the identification of students cross-border. Each issuer has its identity management system to authenticate the identity of students. One example of a national student identifier is based on the Swedish student identity management system (IdM) Ladok\footnote{Ladok Sweden: https://www.student.ladok.se/student/app/studentwebb/}. It provides student identifiers to access EMREX documents for students. Two log-ins are necessary when a Swedish student wants to recognize achievements from an HEI cross-border. One log in Swedish IdM, second in cross-border HEI. When HEI IdM differs in requesting a DC abroad, international students need multiple accounts.

Internationally recognized standards accompany digitization in education. Standards used in education are code lists for countries or courses of study. For example, Swedens' alphanumeric code \textit{752} of ISO 3166-1-alpha-2 or course of study with code \textit{541} Mathematics of ISCED-F 2013. 

Table~\ref{tab1} shows a comparison of external standards used in ELMO and EDS. The first column of Table~\ref{tab1} contains claims used in both ELMO and EDS. The second column \textit{ELMO} contains the external standards' names. The last column \textit{EDS} shows whether a standard is \textit{used} in EDS \cmark \hspace{0.02mm}, \textit{not} \xmark \hspace{0.05mm} or \textit{differs} to ELMO. 

Even more critical than external standards are the digital signatures of DCs. Digital signatures provide secure communication in exchanging DCs in ENs~\cite{kazmirchuk2019digital}. ELMO and EDS use digital signatures to verify the authenticity of digital documents. A comparison of signatures of ELMO and EDS is shown in Table~\ref{tab2}. In ELMO, the authenticity of DCs is verified by an X.509~\cite{chokhani2003internet} certificate of the issuer. EDS supports JWS or JAdES~\cite{ibarz2020bringing} signature of the issuer and holder.
\begin{table}[!t]
\caption{Signatures used in ELMO and EDS}
\begin{center}
\begin{tabular}{|p{2.3cm}|p{2.6cm}|p{2.6cm}|}
\hline
\multirow{2}{*}{\textbf{Claims}} & \multicolumn{2}{c|}{\textbf{Signature Standard}} \\
{} & \textbf{ELMO} & \textbf{EDS}\\
\hline
\textit{Holder} & \hfil (Student) & \hfil (Credential subject) \\
Identifier & Identifier depends on the IdM of the issuer and varies between countries & \hfil DID W3C VC data model \\
Public key location & \hfil \xmark & EBSI ledger DID Doc \\
Signature & \hfil \xmark & \hfil JWS or JAdES \\
\hline
\textit{Issuer} & & \\
Identifier & \hfil National HEI ID & \hfil DID \\
Public key location & \hfil EWP Registry & \hfil EBSI ledger DID Doc \\
Public key standard & \hfil RSA SHA-256 & \hfil EC secp256k1 \\
Signature & \hfil X.509 XML DSig & \hfil JWS or JAdES \\
\hline
\textit{Verifier} & & \\
Verification of DC & Verification of issuers' digital signature. Holders' verification depends on the issuer. & Verification of JWS or JAdES signature of holder and issuer via EBSI ledger. \\
\hline
\end{tabular}
\label{tab2}
\end{center}
\end{table}

The analysis of ELMO and EDS shows differences in data fields (claims) and external standards. For example, ELMO contains the issuers' claim \textit{levelEQF}~\cite{mehaut2012european}, EDS does not as of 6/2022. In EDS, the claim \textit{levelEQF} is available for \textit{qualification}. Another way around, EDS contains an \textit{id} for the holder; ELMO does not by default. A decision for developing ELMO2EDS is to extend EDS by adding missing ELMO claims. Another result of the analysis shows different signature standards. EDS supports blockchain-based signatures, whereas ELMO supports X.509. Therefore, ELMO2EDS omits the XML signature of ELMO and adds a placeholder for EDS Signature in the output data.

The design decisions made for ELMO2EDS result in the following requirements and tasks:
\begin{itemize}
\item \textit{RQ1} Integrity: 
Semantical and syntactical tasks are necessary to integrate ELMO2EDS output data in EDS.
        \begin{itemize}
            \item Mapping structure, claims and external standards of ELMO and EDS
            \item Adding missing ELMO claims in EDS
        \end{itemize}
        
ELMO2EDS adds missing claims of the issuer in EDS. The decision which claims we consider relates to information on german upper secondary school certificates. ELMO and EDS are based on ELM, and the term \textit{missing} applies to a limited extent. Development of EDS considers more and more \textit{missing} claims. EDS is still under development and is not an official standard.

\item \textit{RQ2} Compliance: 
To achieve EBSI compliance of ELMO2EDS, the following \textit{claims and objects} needs to be added to ELMO2EDS to output data (current version adds placeholder):
\begin{itemize}
            \item Holders' and issuers' EBSI DID (e.g., did:ebsi:xyz)
            \item EBSI compliant signature (e.g., JWS)
            \item Schema object. EBSI contains a Trusted Schema Registry (TSR). TSR could be used for external schemas in EDS.
        \end{itemize}
\end{itemize}

The described tasks form the basis of ELMO2EDS. We create templates concluding EDS to achieve \textit{RQ1}. We add placeholders for achieving \textit{RQ2}, as ELMO2EDS is not connected to EBSI yet.

\section{Implementation}
\label{sec:impl}
\begin{figure}[h]
  \centering
  \includegraphics[width=0.4\textwidth]{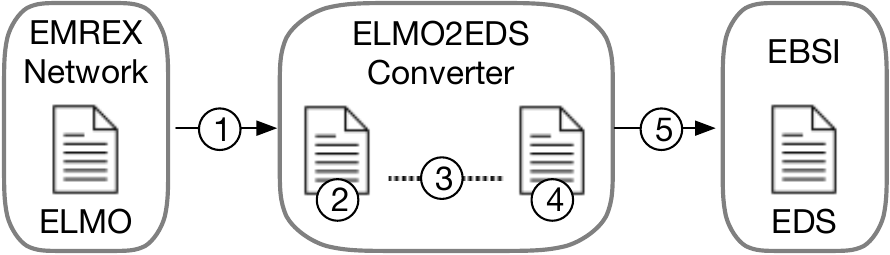}
  \caption{ELMO2EDS: Conversion process from ELMO to EBSI. Mapping and converting process from ELMO to EBSI diploma schema (EDS).}
  \label{fig:ELMO2EDS}
\end{figure}

ELMO2EDS is open-source software accessible via an API to convert data. The converter supports the conversion of upper secondary school certificates and transcripts of records from ELMO to EDS. The process of conversion is shown in Figure~\ref{fig:ELMO2EDS}. In the first step~\ding[1.2]{192} of the conversion, a request is sent to ELMO2EDS to convert an ELMO file to EDS. The request triggers the converter to check whether the requested document is an upper secondary school certificate or transcript of records~\ding[1.2]{193}. The differentiated consideration is required as the EDS structure differs for each type of learning achievement. Depending on the type of ELMO, a template is automatically selected~\ding[1.2]{194} to map claims, external standards, and the structure of ELMO and EDS (\textit{RQ1}). In step~\ding[1.2]{195}, ELMO2EDS adds placeholders for issuers' and holders' signatures (\textit{RQ2}) and returns JSON-LD data format. Transferring converted data to EBSI is shown by process step~\ding[1.2]{196} and is not supported by the converter.

\section{Limitations}
\label{sec:eval}
ELMO2EDS converts ELMO-based credentials into a compliant data format for EBSI. EBSI offers a blockchain-based platform with nodes distributed to all EBP members. With its technology, EBSI aims to improve trust between organizations, governments, and citizens based on the \mbox{Self-Sovereign Identity model~\cite{tan2022blockchain}}. 

The proposed converter acts as a link from EMREX to EBSI, not backward. ELMO2EDS covers parts of SSI Interoperability, defined by Yildiz~\cite{dif_2022}: Semantic data (Layer 4) and credential format (Layer3). ELMO2EDS is limited in usage, as it converts only data. To operate in an SSI environment, ELMO2EDS needs at least a connected SSI agent for issuing and verifying credentials. Furthermore, the SSI-Agent must be compliant with EBSI. Another limitation is the support of different DCs types from EMREX. The converter currently supports only the upper secondary school certificate and transcript of records. 

The comparison of ELMO and EDS shows that some of the standards used by ELMO are currently not in EDS. One ELMO standard not used in EDS in the issuer context is \textit{levelEQF}. The claim \textit{levelEQF} is a candidate for discussion in the future development of EDS. 

Digital signatures can be taken into account during the conversion from ELMO to EDS by adding placeholders. EBSI-compliant signatures can supplement the placeholders of the converted ELMO file. 

\section{Conclusion}
\label{sec:con}

In this paper, we have shown that students are limited in controlling their identities. Students sometimes need two accounts at HEI to request academic achievements. 

The development of cross-border technologies is determined by regulations, especially in Europe. A critical regulation for future ENs is eIDAS, which sets policies for electronic identification and trust services. eIDAS is essential for strengthening the trust in digital transactions between European citizens and organizations. Our proposed converter supports the transition from a server-centric to a user-centric approach and contributes to improving the self-sovereignty of students.

One conclusion is that EBSI deals more flexibly with eIDAS regulation than EMREX. In EBSI, issuers can issue verifiable DC to holders (students), and holders (students) can store verifiable DC in a compliant wallet. Students can present VC for application or recognition of achievements~\cite{olnes2021blockchain}. In EMREX, student data is sent from one HEI to another when students request achievements from abroad. With extensions, EMREX also supports the upload of credits for recognition\footnote{PIM: https://pim-plattform.de/}.

EBSI designed an educational schema related to ELM that can become the standard framework recognized throughout European ENs. It opens opportunities for blockchain-based signatures and personal wallets\footnote{EBSI compliant wallets: https://bit.ly/3ymfZQh}. Probably there will be no Europe-wide standardization of credentials in the education sector. That is why connecting the existing ENs using converters is essential.

We propose how to convert ELMO DC into EDS. ELMO2EDS does not currently meet the requirements of a live system since the following tasks need to be addressed:

\begin{itemize}
    \item Integration of ELMO2EDS into educational institutions
    \item Adding holders' and issuers' DID to credentials by an SSI-Agent
    \item Using EBSI Trusted Registries for including standards (e.g., TIR, TSR)
    \item Cross-border accepted grading schemes (e.g., ECTS for upper secondary school certificates)
    \item Uniform naming of courses of study (e.g., ISCED-F)
\end{itemize}

\section{Outlook}
\label{sec:out}
We presented a solution that addresses a fundamental problem for students: How can students be authenticated in ENs?

At this point, the conversion result of ELMO2EDS is JSON-LD with JWS. There are other potential SSI networks for ELMO2EDS, such as IDunion. IDunion uses privacy-preserving CL signatures instead of the only supported signature of ELMO2EDS, JWS. Integrating ELMO2EDS into IDunion is a potential research topic that allows the converter to run as a middleware.

Education is in a digital transformation. Educational credentials are increasingly being exchanged and verified in ENs. E-learning platforms are already taking place at the elementary school level.

Self-Sovereign Identity promises users more control over their data, especially when handling sensitive data online. Since DCs contain personal information, these must be specially protected, not least because of regulations \mbox{such as eIDAS}. 

\section*{Acknowledgment}
We want to thank Leon Rohmann for his excellent work on the project. Furthermore, we thank Adrian Michalke, Benjamin Burde, Guido Bacharach, and Helmut Nehrenheim.

\bibliographystyle{IEEEtran}
\bibliography{bibliography.bib}

\end{document}